\documentclass{PoS}

\title{The Be X-ray Binary Outburst Zoo}

\ShortTitle{The Be X-ray Binary Outburst Zoo}

\author{\speaker{Peter Kretschmar}\\ 
        European Space Astronomy Centre (ESA/ESAC), Science Operations Department, 
        Villanueva de la Ca{\~n}ada (Madrid), Spain\\
        E-mail: \email{peter.kretschmar@esa.int}}

\author{Elisa Nespoli\\
        Observatorio Astron\'omico de la Universidad de Valencia, Spain}

\author{Pablo Reig\\
        IESL, Foundation for Research and Technology-Hellas, Heraklion, Greece\\
        Institute of Theoretical \& Computational Physics, University of Crete, Greece}

\author{Friedrich Anders\\
        Leibniz Institute for Astrophysics Potsdam, Germany}

\abstract{Be X-ray binaries are among the best known transient high-energy sources. Their outbursts are commonly classified into a simple scheme of 'normal' and 'giant' outbursts, but a closer look shows that actual outbursts do not always follow this simple scheme. Recent data show a variety of properties, like pre-flares, shifts of the outburst peaks with respect to the periastron, multi-peaked outbursts etc. We present results from a systematic study of a large number of outbursts monitored by various space missions, comparing outburst properties and their relation to system parameters and current theoretical understanding.}

\FullConference{"An INTEGRAL view of the high-energy sky (the first 10 years)"
9th INTEGRAL Workshop and celebration of the 10th anniversary of the launch,\\
		October 15-19, 2012\\
		Bibliotheque Nationale de France, Paris, France}

\begin{document}

\section{Be X-ray Binaries}

Be~X-ray binaries (BeXRBs) are a subset of the class of High-Mass X-ray Binaries (HMXBs).
In these binaries the mass-donor is a Be star, i.e., an O or B star which is not in the
supergiant stage and from which at some point spectral lines in emission have been observed.
These stars also show increased infrared emission, when compared to non-emitting stars of the
same type. Both phenomena are explained by an extended circumstellar disk,
which is formed in a way that is not completely understood, although fast rotation of the
star plays a role, and pulsations are probably involved \cite{PorterRivinius:2003}. 
The compact object in BeXRBs is typically a neutron star, usually an accreting pulsar; no Be star / Black Hole binary is known so far. For a recent review see \cite{Reig:2011Review}.

BeXRBs are very variable on many timescales and at different wavelengths. In the X-rays,
most BeXRB are transient systems, mostly observable during outbursts, when these sources 
can be among the brightest sources in the X-ray sky for a few weeks. Historically these 
outbursts have been usually classified into two types: Type-I or normal outbursts are
connected to the periastron passage of the neutron star and less bright; type~II or giant
outbursts are not clearly related to orbital phase and can reach luminosities up to the
Eddington luminosity for a neutron star. At least during some outbursts, the presence of
quasi-periodic oscillations (QPOs) indicates formation of an accretion disk in the outburst
\cite{HayasakiOkazaki:2004}.

There is a complex connection between the long-term evolution of the system brightness and
emission line measurements and the X-ray activity. For some cases these variations can be
explained by a build up and disruption of the circumstellar disk 
(see \cite{Reig:2011Review} and references therein).

The different forms of activity in BeXRBs, from sometimes years of quiescence to semi-regular 
outbursts and giant outbursts has been explained with some success by the viscous decretion disc model \cite[and related publications]{OkazakiNegueruela:2001}. 
A recent publication \cite{OkazakiHayasakiMoritani:2012} proposes a new scenario in which
normal outbursts are fed by radiatively inefficient accretion flows while giant
outbursts are fed by Bondi-Hoyle-Lyttleton, respectively, from a circumstellar disk, 
misaligned with the orbital plane.

Despite these modeling efforts, the observed outbursts in BeXRBs still show a large variety of
properties, that are not easily explained within the current framework. First of all, in the
range of variations there are occasional `intermediate' outbursts, which cannot be easily 
classified into the type~I/II scheme. There is a large variety of shapes and durations,
including pre-cursor peaks and double-peaked outbursts (see, e.g., \cite{Caballero:2010IWS9}
and \cite{ReigNespoli:2013a}).
For multiple consecutive outbursts of 1A\,0535+262 a shifting 
outburst orbital phase, or alternatively an activity period of $\sim$115\,d instead of the
110.2\,d orbital period was observed \cite{{Nakajima:2010fym20}}.

The fact that BeXRBs are readily observed over a wide range of luminosities makes
them interesting `laboratories' for the physics of accretion and the accretion column.
The pulse period changes observed during the outbursts (see \cite{Reig:2011Review} and
references therein) linked to the brightness evolution yield information about
the interaction of the accreted matter with the neutron star's magnetic field. 
More information can be obtained from QPOs, when they are observed. Changes in 
spectral parameters and pulse profiles \cite{ReigNespoli:2013a,Caballero:2011}
link to the accretion column structure and the emission processes in the column.

\begin{figure} 
\centerline{\includegraphics[width=.86\textwidth]{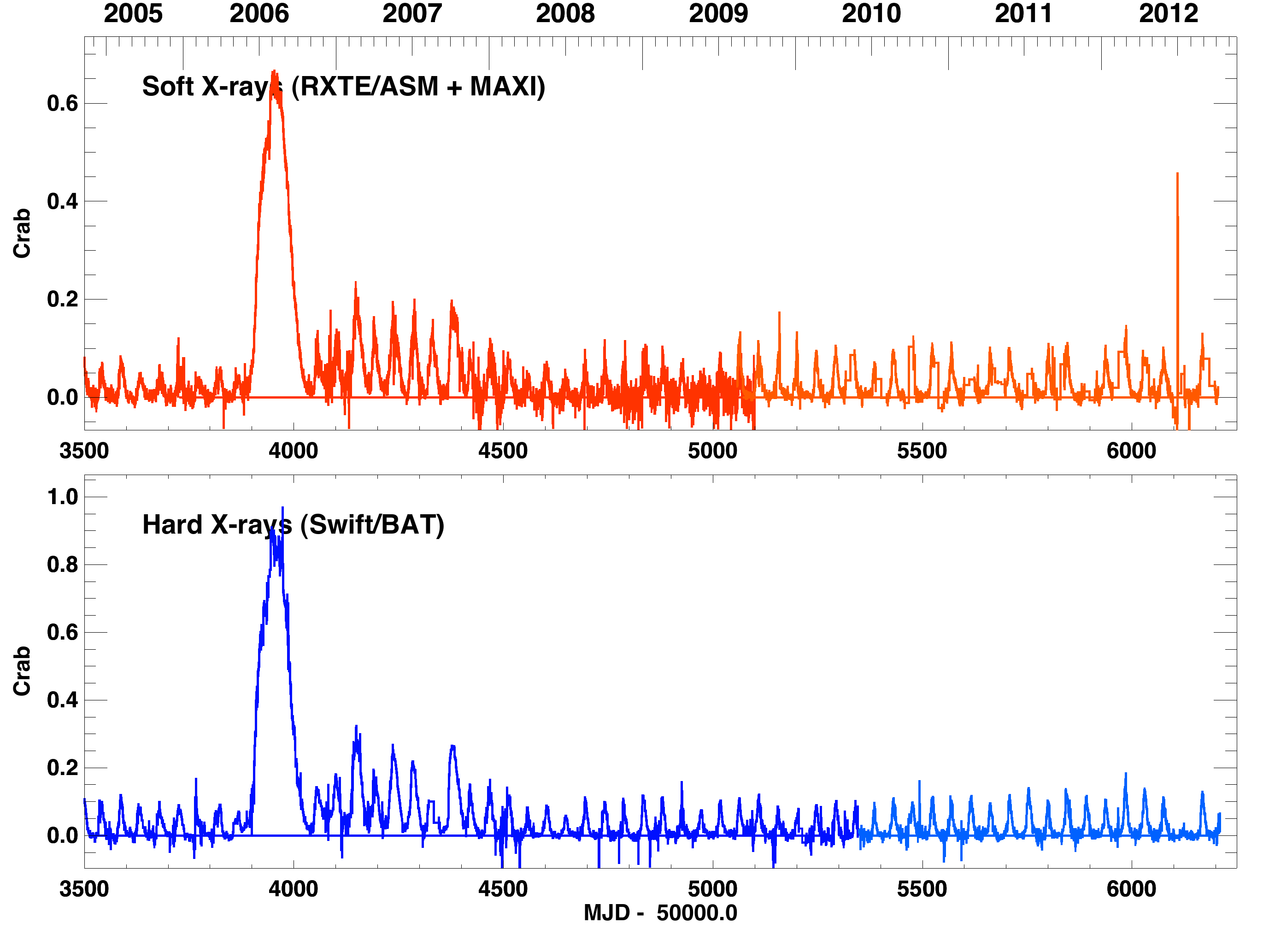}}
\caption{Long-term lightcurve in Crab units of EXO\,2030+375. The giant outburst of 2006 spanned multiple orbits, while
the regular outbursts are spaced by the orbital period (46\,d).} \label{fig:exo2030a} 
\end{figure}
\begin{figure} 
\centerline{\includegraphics[width=.86\textwidth]{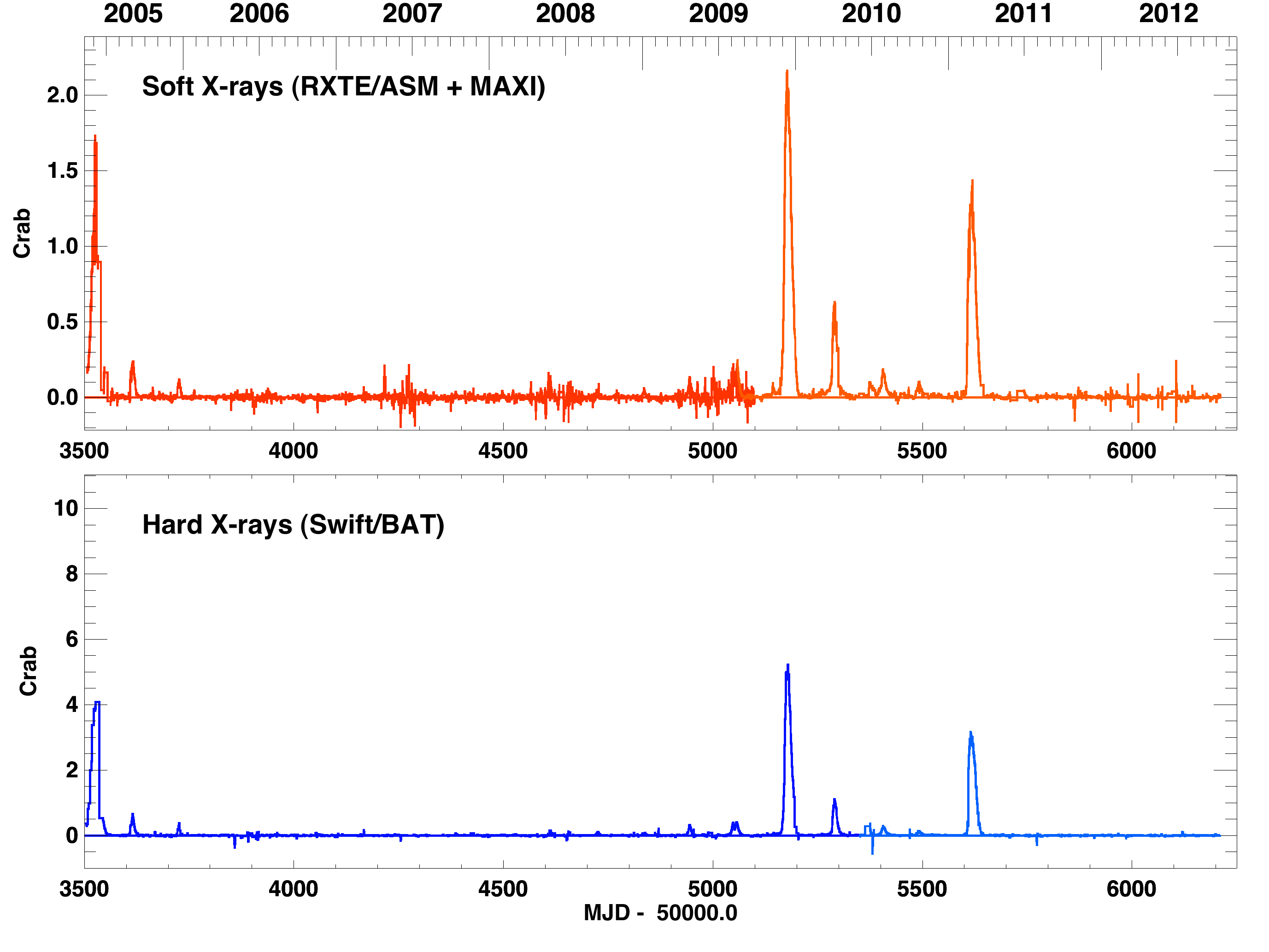}}
\caption{Long-term lightcurve in Crab units of 1A\,0535+262,. In this source the giant outbursts cover only a fraction of the orbital period (110\,d).} \label{fig:a0535} 
\end{figure}

\begin{figure} 
\makebox[0.95\textwidth]{%
\includegraphics[width=.42\textwidth]{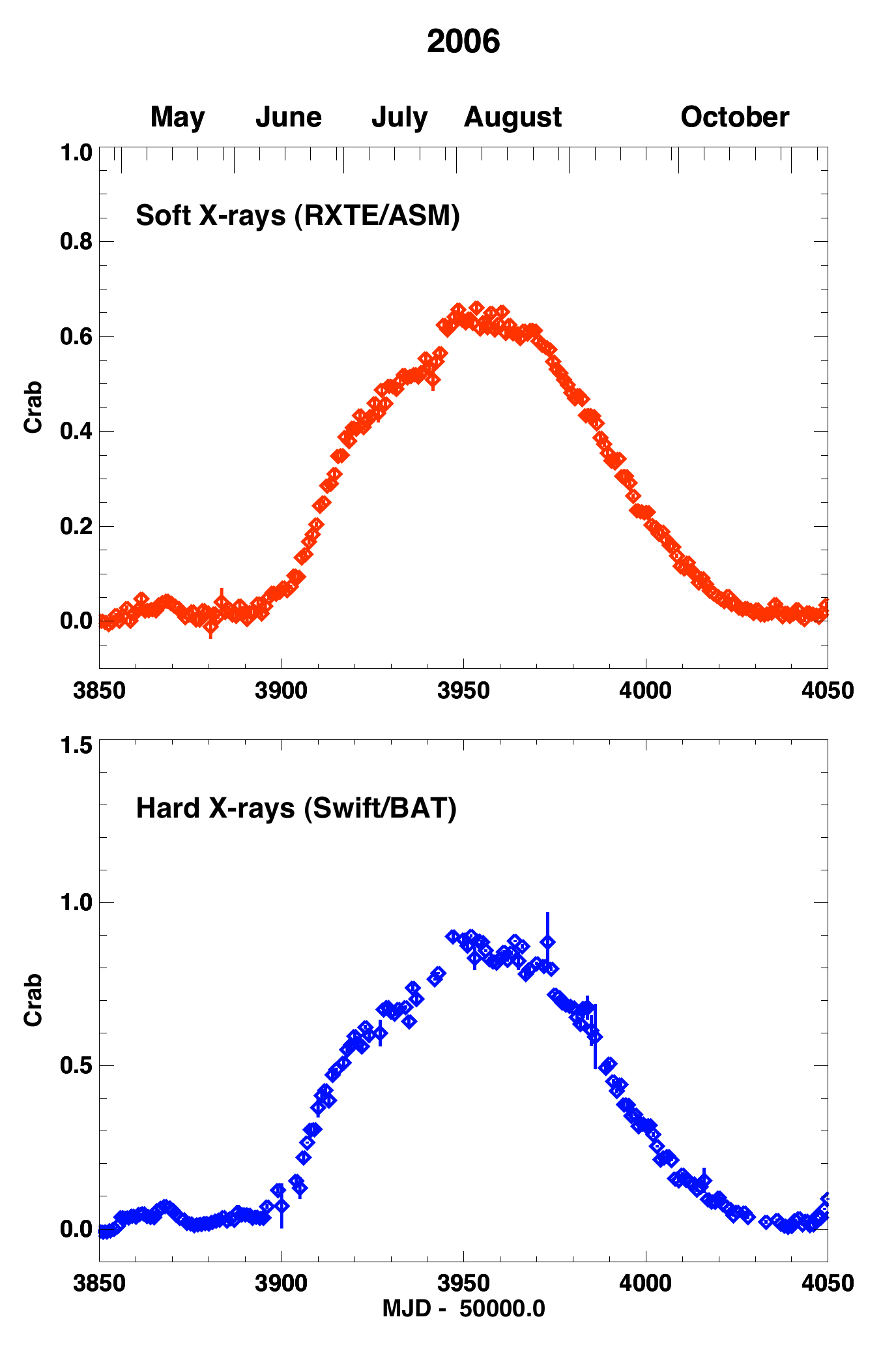}\hfil
\includegraphics[width=.42\textwidth]{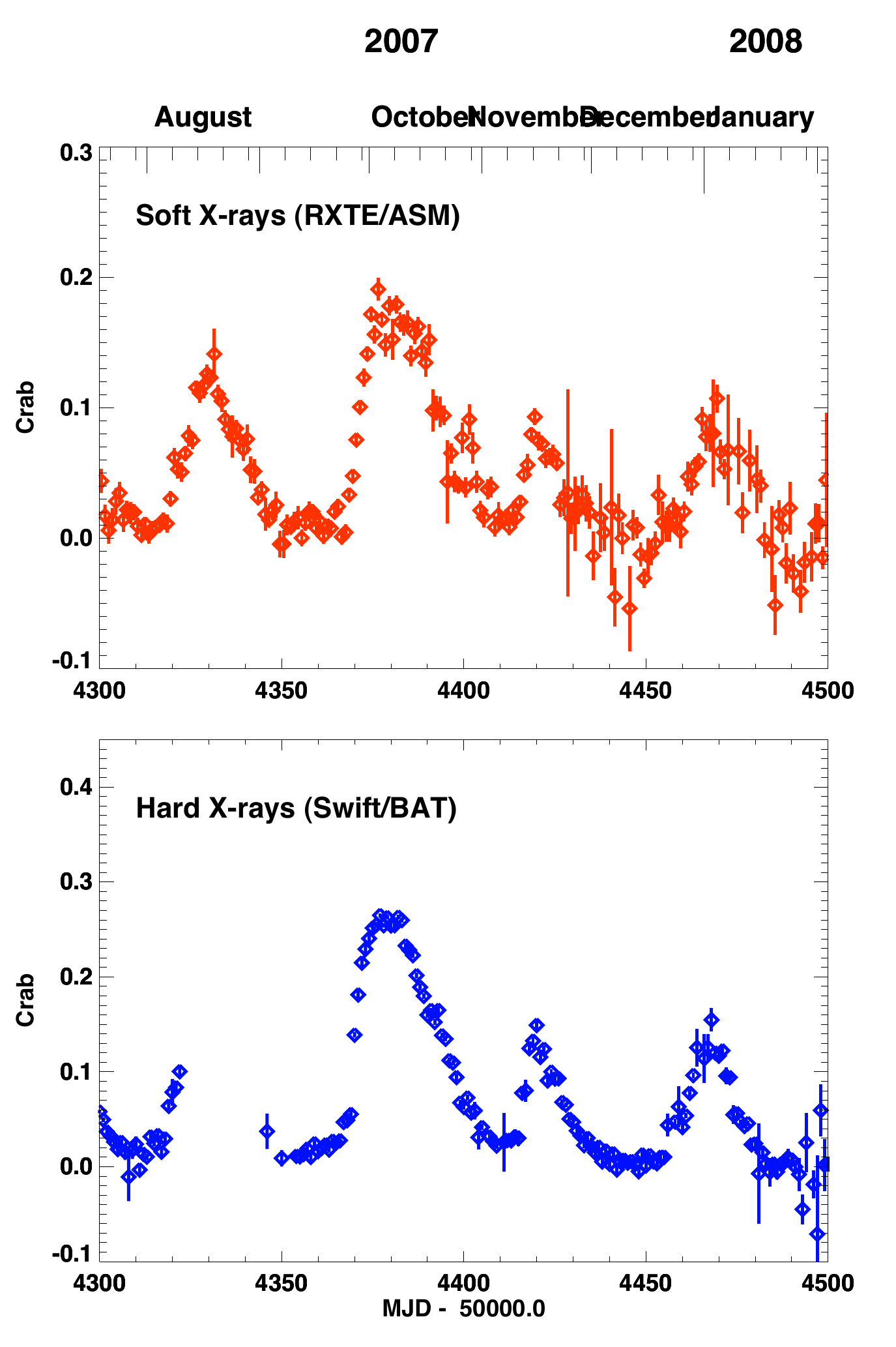}}
\caption{Comparison of a giant outburst of EXO\,2030+375 with a sequence of regular normal outbursts, typical for this source. Both graphs cover the same time scale, but have different
Y axis scales. Note the varying shapes of the normal outbursts.} \label{fig:exo2030b} 
\end{figure}

\begin{figure} 
\makebox[0.95\textwidth]{%
\includegraphics[width=.42\textwidth]{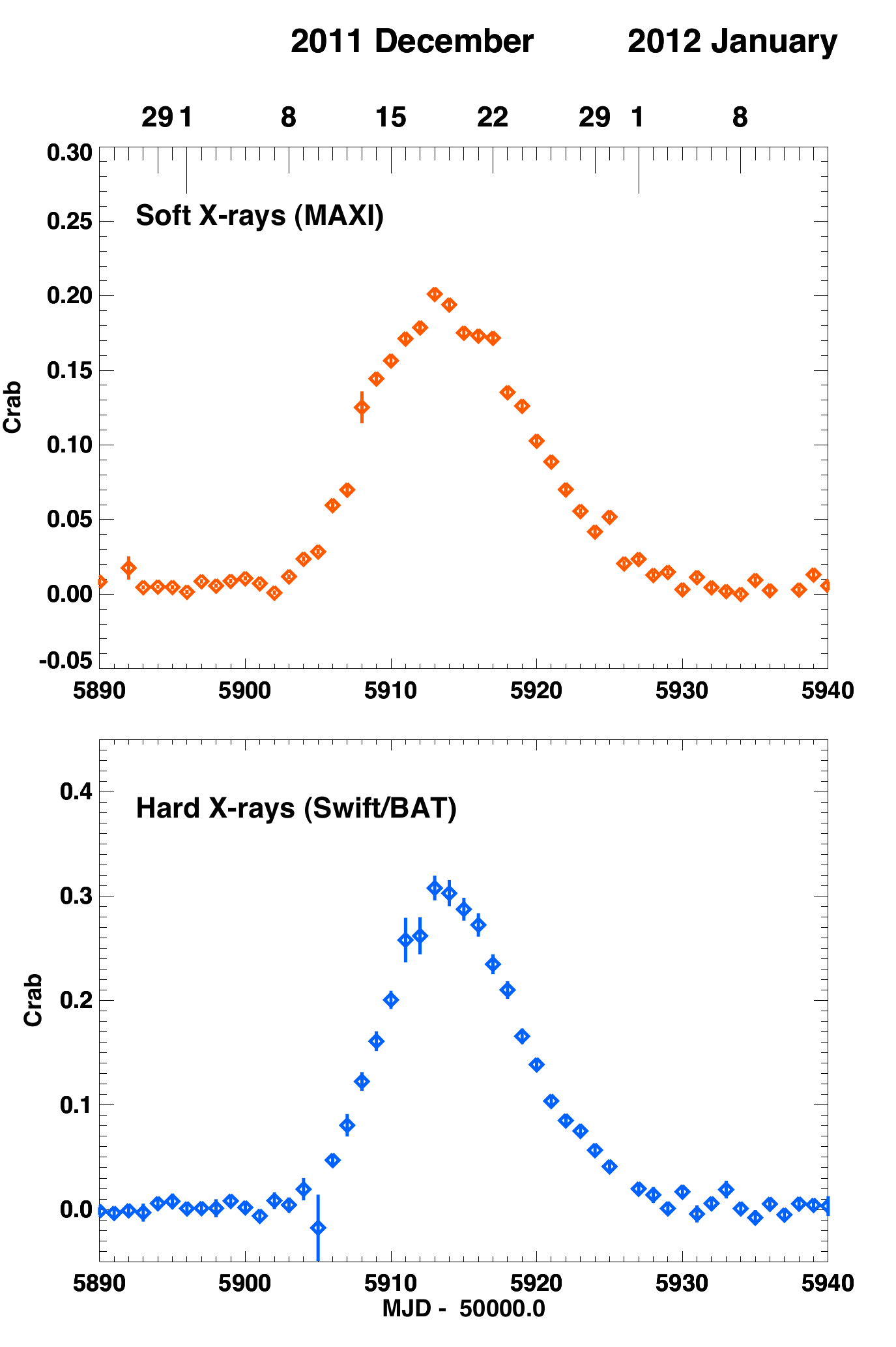}\hfil
\includegraphics[width=.42\textwidth]{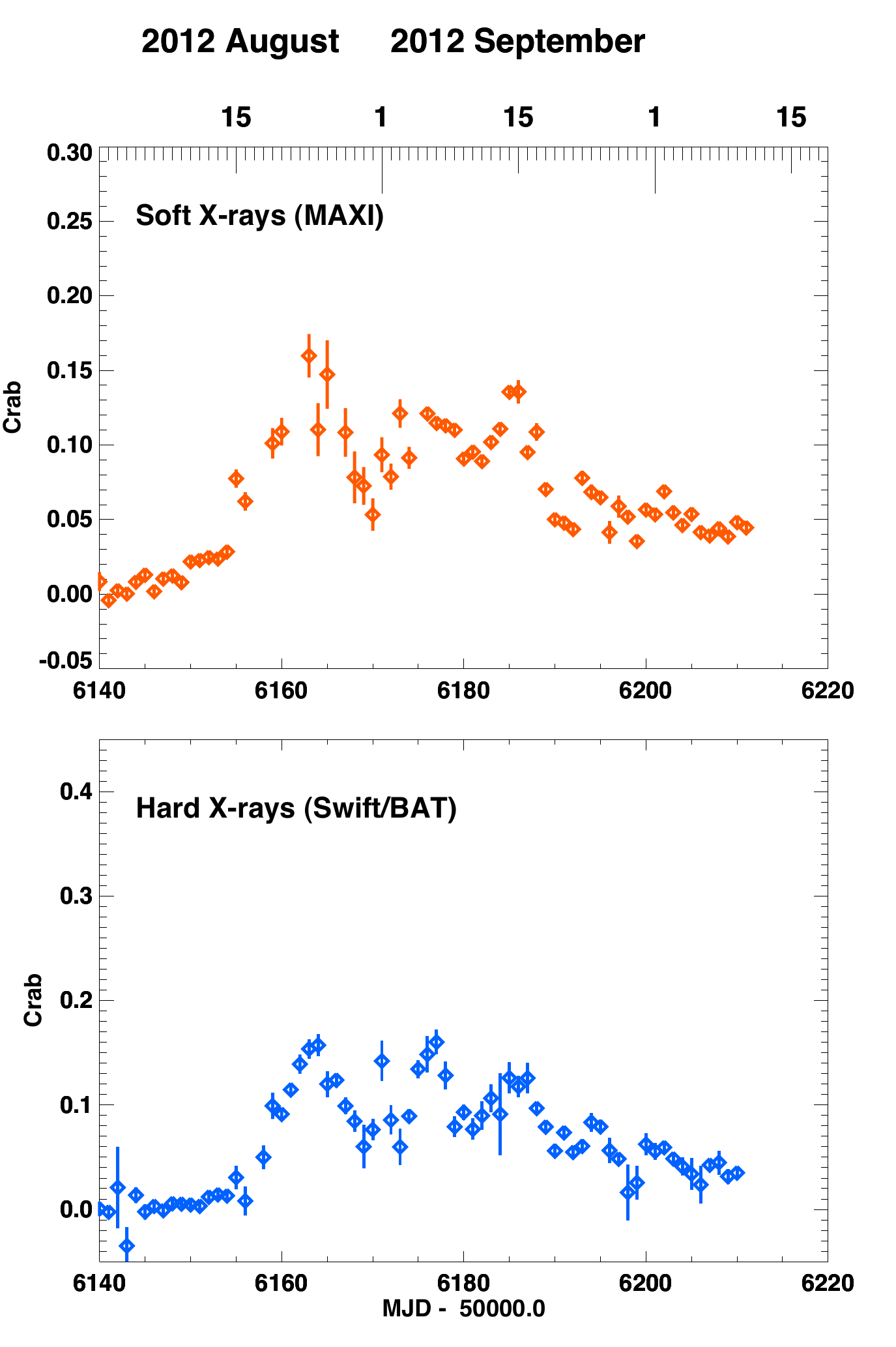}}
\caption{Comparison of two subsequent outbursts of GRO\,J1008$-$57 (orbital period: 248\,d). Note the different time scales covered. The second outburst was still ongoing when the graph was made.}
\label{fig:groj1008} 
\end{figure}

\section{A Systematic Study of BeXRB Outbursts}
In order to allow a more general comparison across different sources, we have begun a 
systematic study of BeXRB outbursts using the available
data from all-sky monitoring instruments, i.e., \textit{CGRO/BATSE}, \textit{RXTE/ASM},
\textit{Swift/BAT} and \textit{MAXI}. The idea is to derive a consistent catalogue of
BeXRB outbursts seen by these monitors with outburst times, orbital phases, durations, 
fluxes, shapes, etc., and to
compare with the known parameters of these systems.

The systematic compilation and analysis of outbursts is still ongoing, but a few preliminary conclusion can be drawn from a subset involving 17 sources and about 170 transient events:
\begin{itemize}
\item The outburst shapes are very variable with no evident dependence on the source 
      or on the parameters of the specific outbursts. 
      The `textbook case' of a relative fast rise and a slower decay is seen in less
      than 1/3 of the outbursts studied so far. Especially fainter outbursts show a
      large variety of shapes.
\item There is a significant fraction ($\sim$1/4) of outbursts with complex shapes,
      e.g., multiple peaks, partial rise, flares, etc.
\item There is no evident correlation between the length of outbursts and the orbital
      period of the system. Most outbursts take place in less than half of the orbital
      period, but some giant outbursts have continued for up to $\sim$3 orbits.
      In absolute terms, most outbursts are shorter than a month, but there is a long
      tail of a few longer outbursts, going on for several months.          
\end{itemize}

\section{Discussion \& Outlook}
At the moment, efforts to model BeXRB outbursts or to trace accretion physics have been 
largely based on a few sources and/or specific outbursts. With our outburst catalogue 
in preparation we hope 
to incite more systematic comparisons between theory and observations, leading in the
long run to updated, more realistic models for these interesting systems.

\end{document}